\documentclass{article} 
\usepackage{epsfig}

\begin{document}

\title{Level density and thermal properties in rare earth nuclei}
\author{A.~Schiller\footnote{E-mail address: Andreas.Schiller@fys.uio.no}, 
M.~Guttormsen, M.~Hjorth-Jensen, E.~Melby,\\ 
J.~Rekstad, and S.~Siem,\\ 
Department of Physics, University of Oslo,\\  
P.O.Box 1048 Blindern, N-0316 Oslo 3, Norway}

\date{}

\maketitle

\begin{abstract}
A convergent method to extract the nuclear level density and the $\gamma$-ray
strength function from primary $\gamma$-ray spectra has been established. 
Thermodynamical quantities have been obtained within the microcanonical and
canonical ensemble theory. Structures in the caloric curve and in the heat
capacity curve are interpreted as fingerprints of breaking of Cooper pairs and 
quenching of pairing correlations. The strength function can be described 
using models and common parameterizations for the E1, M1 and pygmy resonance 
strength. However, a significant decrease of the pygmy resonance strength at 
finite temperatures has been observed.
\end{abstract}

\section{Introduction}

Investigation of nuclear level density is an old problem in nuclear physics. 
The first theoretical attempt to describe nuclear level density was done by 
Bethe in 1936 \cite{Be36}. In order to do so, he introduced thermodynamical 
quantities like temperature and entropy, showing how closely related nuclear 
level density and thermodynamics in nuclei are. With the discovery of pairing 
correlations, their effect on nuclear level density, temperature and heat 
capacity has been explored early in schematic calculations \cite{SY63}. Today, 
the Monte-Carlo shell model technique \cite{KD97} can estimate nuclear level 
density \cite{NA97} reliably for heavy mid-shell nuclei like dysprosium 
\cite{WK98}. 

On the experimental side, the main sources of information on nuclear level 
density have been counting of discrete levels in the vicinity the ground state 
(see e.g.\ \cite{FS96}) and neutron resonance spacing data (see e.g.\ 
\cite{IM92}). Recently, the Oslo group has reported on a new method to extract
level density and $\gamma$-ray strength function from primary $\gamma$-ray
spectra \cite{SB00}. 

Important applications of nuclear level densities are Hauser-Feshbach type of
calculations \cite{HF52} of nuclear reaction cross-sections. These reaction
cross-sections are important input parameters in large network calculations of
stellar evolution \cite{RT97}. The reaction cross-sections can also be used to
estimate the efficiency of accelerator-driven transmutation of nuclear waste.

Also radiative strength functions have been examined since long time. The first
estimate of $\gamma$-ray strength functions within the single-particle shell 
model was done by Weisskopf in 1951 \cite{We51}. However, this model of 
energy-independent strength functions failed particularly badly with E1 
transitions. First some ten years later \cite{Ax62}, experimental data on 
electric dipole transitions over a large energy range could be explained 
consistently within one model. Today, refined schematic models of the giant 
dipole resonance, taking into account temperature dependence, are available
\cite{KM83,Si86}, while low-lying dipole strength can be reliably estimated 
within microscopic random-phase approximation calculations for rare earth 
nuclei \cite{SS97a,SS97b}.

Experimentally, the total radiative strength function can be measured by 
absorption methods \cite{GL81}. At energies below the neutron separation energy
it can be estimated from radiative neutron capture, usually assuming a model 
for the nuclear level density. These experiments involve either the total
$\gamma$-ray spectrum \cite{IK86} or two-step $\gamma$ cascades \cite{BC95} 
(see also the talk of A. M. Sukhovoj in this Volume). Our newly developed 
method \cite{SB00} gives now for the first time the opportunity to extract
level density and radiative strength function simultaneously without assuming
any model for either of them.

Applications of radiative strength functions can again be found in nuclear
astrophysics. Especially the existence of a soft dipole mode in neutron rich 
nuclei can have a large impact on the $(n,\gamma)$ reaction rates of r-process 
nuclei \cite{Go99}.

In Sect.\ \ref{sec:2}, we discuss the experimental details and the main 
assumptions of our data analysis method. In Sect.\ \ref{sec:3}, results for the
level density and thermodynamical quantities are shown. In Sect.\ \ref{sec:4}, 
the radiative strength function is discussed, and we conclude the talk in 
Sect.\ \ref{sec:5}.

\section{Experimental details and data analysis}
\label{sec:2}

The experiments were carried out at the Oslo Cyclotron Laboratory at the 
University of Oslo, using an MC35 Scanditronix cyclotron with a $^3$He beam 
energy of 45~MeV and a beam intensity of typically 1~nA. The experiments were 
usually running for two weeks. The targets consist of self-supporting, 
isotopically enriched ($\sim$95\%) metal foils of $\sim$2.0~mg/cm$^2$ 
thickness, glued on an aluminum frame. Particle identification and energy 
measurements were performed by a ring of 8~Si(Li) particle telescopes mounted 
at 45$^\circ$ with respect to the beam axis. The telescopes consist of a front 
and end detector with thicknesses of some 150 and 3000~$\mu$m respectively and 
can effectively stop $\alpha$ particles with energies up to 60~MeV. The 
$\gamma$ rays were detected by an array of 28~5''$\times$5''~NaI(Tl) detectors 
(CACTUS) \cite{GA90} covering a solid angle of $\sim$15\% of $4\pi$. Three 60\%
Ge(HP) detectors were used to monitor the selectivity of the reaction and the 
entrance spin distribution of the product nuclei. During one experimental run, 
data can be recorded and sorted out simultaneously for the ($^3$He,$^3$He') and
the ($^3$He,$\alpha$) reaction on the same target.

In the data analysis, the ejectile energy can be transformed into excitation
energy of the product nucleus, since the reaction kinematic is uniquely
determined. In the next step, the $\gamma$-ray spectra are unfolded 
\cite{GT96}, using measured response functions of the CACTUS detector array.
Afterwards, the primary $\gamma$-ray spectra can be extracted, using the 
subtraction technique of Ref.~\cite{GR87}. In order to be able to apply this 
technique, the entrance point in excitation energy of the product nucleus has
to be known and all excitation energies up to a certain limit have to be 
scanned in the experiment. The basic assumption behind the first-generation 
method is that the $\gamma$-ray spectrum of any excitation energy bin is 
independent of the way how states in this bin are populated (e.g.\ direct
population by a nuclear reaction, or population by the same nuclear reaction at
some higher entrance energy and followed by one or several subsequent $\gamma$ 
rays). This assumption is not completely valid at low excitation energies where
$\gamma$ decay competes effectively with thermalization processes and the 
nuclear reactions applied exhibit a more direct than compound character. Also 
possibly different spin and parity distributions of levels populated at 
different excitation energies by the same nuclear reaction can violate this 
assumption. However, in a recent investigation of this matter, we could not 
find any severe problems with the first-generation method \cite{SG00}. 

\begin{figure}[htb!]\centering
\mbox{\epsfig{figure=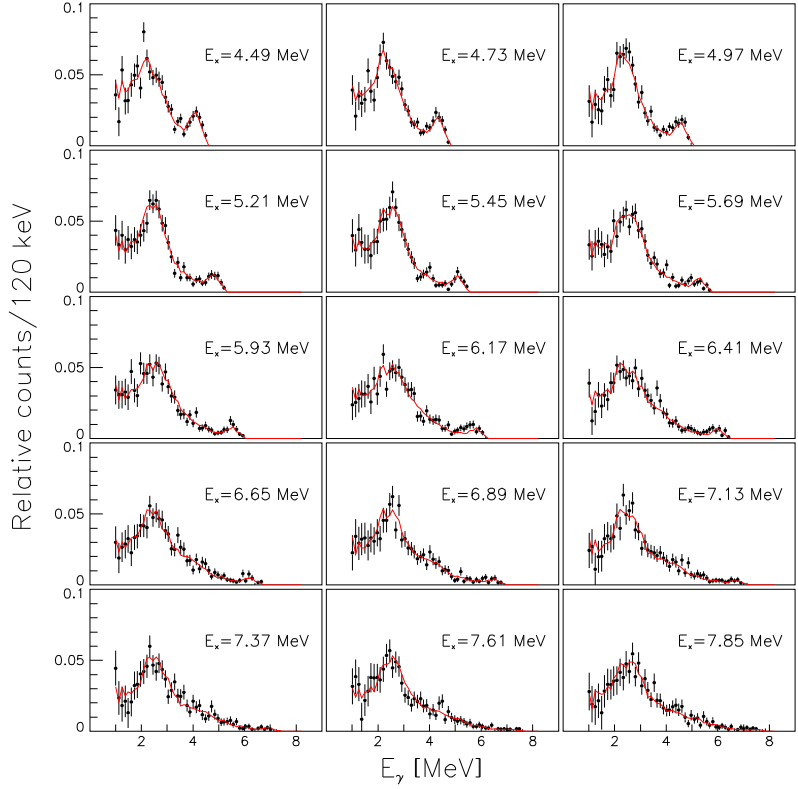,height=12.1cm}}
\caption{Normalized experimental primary $\gamma$-ray spectra with estimated
errors (data points) and fit using the factorization of Eq.~(\ref{eq:1}) 
(solid lines). The data are taken from the $^{162}$Dy($^3$He,$^3$He')$^{162}$Dy
reaction.}
\label{fig:1}
\end{figure}

The primary $\gamma$-ray spectra (see Fig.~\ref{fig:1}) are the starting point 
of the discussion in this talk. According to the Brink-Axel hypothesis 
\cite{Br55,Ax62}, the primary $\gamma$-ray matrix can be factorized into two 
functions of one variables using
\begin{equation}
\label{eq:1}
\Gamma(E,E_\gamma)\propto\rho(E-E_\gamma)\,F(E_\gamma),
\end{equation}
where $\rho$ is the level density and $F$ is a $\gamma$-ray energy-dependent
factor, proportional to the total radiative strength function i.e.\ 
\begin{equation} 
F(E_\gamma)\propto\sum_{XL}E_\gamma^{(2L+1)}\,f_{XL}(E_\gamma).
\end{equation}
In Eq.~(\ref{eq:1}), a temperature-independent radiative strength function $f$
is assumed. Today, we know that at least the E1 strength function is 
temperature dependent, a fact which has already been incorporated in several 
models \cite{KM83,Si86}. However, we found for our data that the factorization 
according to Eq.~(\ref{eq:1}) works remarkably well (see Fig.~\ref{fig:1}) 
which indicates that for low and slowly varying temperatures as in our case, 
the Brink-Axel hypothesis is approximately valid. 

The details of the method to extract level density and radiative strength 
function from primary $\gamma$-ray spectra can be found in \cite{SB00}. An 
extension of this method to temperature-dependent radiative strength function 
is discussed in Sect.~\ref{sec:4}. One detail of the method should be mentioned
here. The method does not yield absolute values of the level density and the 
radiative strength function. Also the slope of these two functions is 
undetermined. Actually, all functions $\tilde{\rho}$ and $\tilde{F}$ obtained 
by the transformation 
\begin{eqnarray}
\tilde{\rho}(E-E_\gamma)&=&A\,\exp(\alpha\,[E-E_\gamma])\,
\rho(E-E_\gamma)\\
\tilde{F}(E_\gamma)&=&B\,\exp(\alpha\,E_\gamma)\,F(E_\gamma)
\end{eqnarray}
of any particular solution $(\rho,F)$ will fit our primary $\gamma$-ray matrix 
equally, since the area of the first-generation spectrum are normalized to  
unity for every excitation energy bin $E$, i.e. 
\begin{equation}
\Gamma(E,E_\gamma)=\frac{\rho(E-E_\gamma)\,F(E_\gamma)}
{\sum_{E_\gamma}\rho(E-E_\gamma)\,F(E_\gamma)}.
\end{equation}
In order to determine the parameters $A$ and $\alpha$, i.e.\ the absolute value
and the slope of the level density, we fit our extracted level density curve to
the known number of discrete levels in the vicinity of the ground state 
\cite{FS96} and to the level density estimate obtained from neutron resonance 
spacing data \cite{IM92} at the neutron binding energy. The only remaining free
parameter then is the absolute value of the $\gamma$-ray energy-dependent 
factor $F$, which can be determined from the average total radiative widths of 
neutron capture resonances \cite{Mu84} by
\begin{equation}
\langle\Gamma_\gamma(E,I,\Pi)\rangle=\frac{1}{\rho(E,I,\Pi)}
\sum_{XL}\sum_{I_f,\Pi_f}\int_{E_\gamma=0}^{E}{\mathrm{d}}E_\gamma
E_\gamma^{2L+1}f_{XL}(E_\gamma)\,\rho(E-E_\gamma,I_f,\Pi_f)
\end{equation}
(see e.g.\ \cite{KU90}).

In this talk, we will discuss level density and radiative strength function 
of $^{161,162}$Dy and $^{171,172}$Yb obtained from ($^3$He,$\alpha$) reaction 
data and radiative strength function of $^{162}$Dy obtained from 
($^3$He,$^3$He') reaction data.

\section{Level density and thermodynamical quantities}
\label{sec:3}

\begin{figure}[htb!]\centering
\mbox{\epsfig{figure=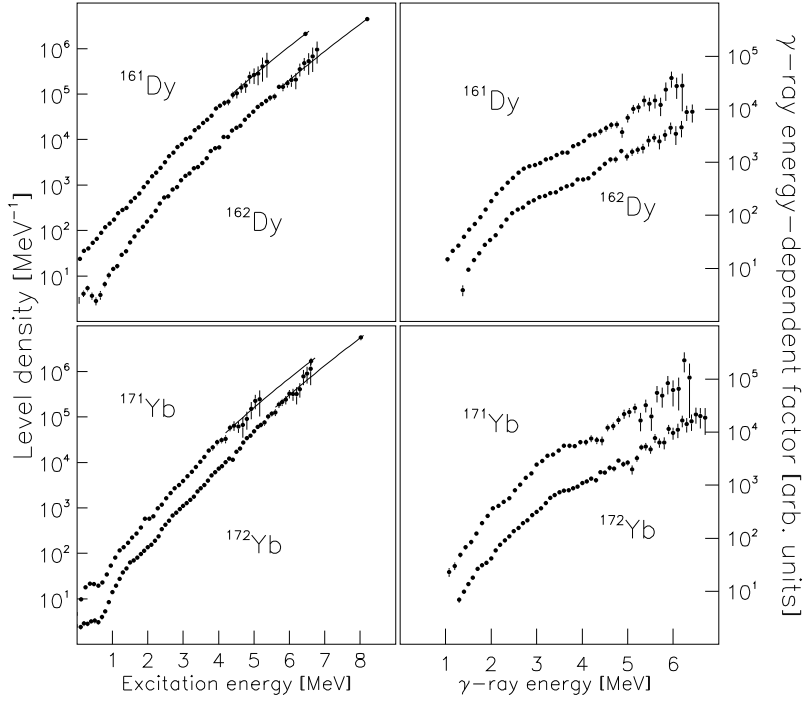,height=12.1cm}}
\caption{Level density and $\gamma$-ray energy-dependent factor $F(E_\gamma)$ 
of $^{161,162}$Dy and $^{171,172}$Yb from ($^3$He,$\alpha$) reaction data. The 
error bars show the experimental uncertainties. The solid lines are 
extrapolations based on a shifted Fermi-gas model. The isolated points at the 
neutron binding energy were obtained from neutron resonance spacing data.}
\label{fig:2}
\end{figure}

In Fig.~\ref{fig:2}, the nuclear level density and the $\gamma$-ray 
energy-dependent factor $F(E_\gamma)$ for the nuclei $^{161,162}$Dy and 
$^{171,172}$Yb are shown. In this section, we will mainly discuss the physics 
of the nuclear level density. First of all, the experimental curves can be
compared to popular parameterizations of the nuclear level density, like those 
of Gilbert and Cameron \cite{GC65} or of von Egidy et al.\ \cite{ES87}. This 
has been done in \cite{GH00} and the conclusion is that neither of the two
parameterizations can describe our data well. However, the data favor the 
concept of a composite level density formula as proposed in \cite{GC65} with a
constant-temperature level density part from above 1--2~MeV and up to 
approximately the neutron binding energy $B_n$. Another important aspect is 
that the experimental data of the odd and even nuclei show a relative shift in 
the order of the effective pairing energy 
$\Delta^{\mathit{eff}}(N,Z)=\Delta_p(N,Z)+\Delta_n(N,Z)-\Delta_p(N-1,Z)$
\cite{GH00}, thus the data support the concept of shifted level density 
formulas.

From level densities, one can easily calculate thermodynamical quantities like
entropy $S$, temperature $T$, heat capacity $C_V$, the canonical partition
function $Z$ and the average excitation energy in the canonical ensemble 
$\langle E\rangle$. 
Within the microcanonical ensemble, one obtains (in this work $k_B$=1)
\begin{eqnarray}
S(E)&=&\ln\rho(E)+S_0\\
\label{eq:8}
T(E)&=&\left(\frac{\partial S(E)}{\partial E}\right)_V^{-1}\\
C_V(E)&=&\left(\frac{\partial T(E)}{\partial E}\right)_V^{-1}
\end{eqnarray}
and in the canonical ensemble, one gets
\begin{eqnarray}
\label{eq:10}
Z(T)&=&\int_0^\infty N\rho(E)\exp(-E/T){\mathrm{d}}E\\
S(T)&=&\frac{\partial}{\partial T}[T\ln Z(T)]\\
\langle E(T)\rangle&=&T^2\frac{\partial}{\partial T}\ln Z(T)\\
C_V(T)&=&\left(\frac{\partial\langle E(T)\rangle}{\partial T}\right)_V.
\end{eqnarray}
The quantities $S_0$ and $N$ are necessary, since the level density is only 
proportional to the energy surface in the phase space $W$.

\begin{figure}[htb!]\centering
\mbox{\epsfig{figure=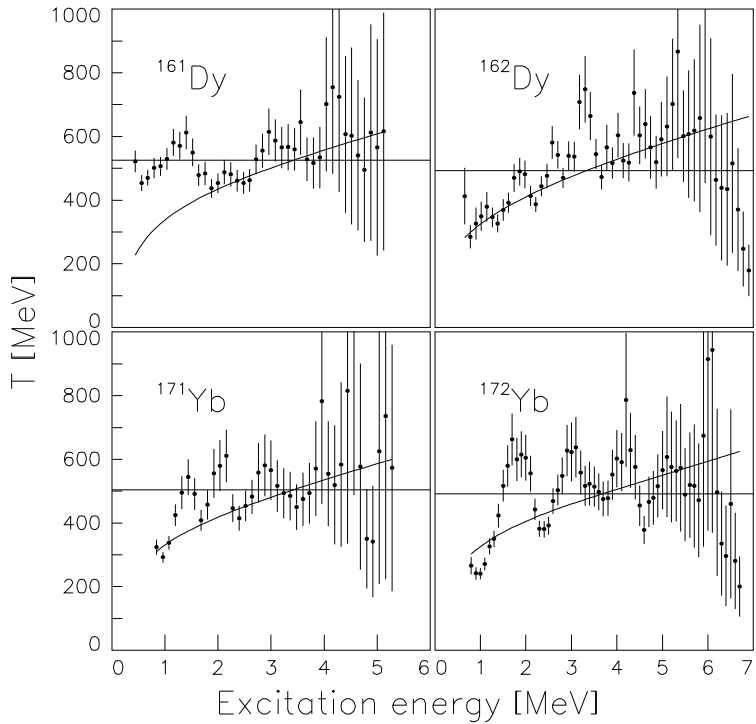,height=12.1cm}}
\caption{Caloric curve of $^{161,162}$Dy and $^{171,172}$Yb in the 
microcanonical ensemble (data points) and the canonical ensemble (line). The
straight line indicates the critical temperature $T_c$.}
\label{fig:3}
\end{figure}

In principle, one should only consider the microcanonical ensemble, since the
nucleus is a closed system. However, the canonical and even the grand-canonical
ensemble have often been used \cite{Be36,KD97} to describe thermodynamical 
properties of nuclei. In \cite{GB00}, the microcanonical and canonical entropy
is discussed and compared to a simple model. One result of this discussion is
that the small bumps in the experimental level density curves (see 
Fig.~\ref{fig:2}) can be interpreted in terms of breaking of Cooper pairs. 
These bumps can even be enhanced by derivating, see Eq.~(\ref{eq:8}), yielding 
the experimental caloric curve in the microcanonical ensemble (see data points 
in Fig.~\ref{fig:3} and discussion in \cite{MB99}). Another important result is
that the entropy excess of the odd nuclei relative to the even nuclei can be 
used to calculate the entropy of one quasiparticle. It is surprising that the 
quasiparticle entropy is constant $1.70(15)\,k_B$ over the whole excitation 
energy region investigated in \cite{GB00}.

When calculating the partition function in the canonical ensemble (see Eq.\ 
(\ref{eq:10})), a strong smoothing is introduced due to the Laplace
transformation involved. It is also worth noting that in order to calculate 
thermodynamical quantities reliably up to $T\sim 1$~MeV, one has to know the 
level density up to $\sim 40$~MeV. Since the experimental level density curves
are only known up close to the neutron binding energy, they had to be 
extrapolated by a model. We have chosen the shifted Fermi-gas parameterization 
of von Egidy et al.\ \cite{ES87} multiplied by a constant factor in order to 
match the neutron resonance spacing data.

\begin{figure}[htb!]\centering
\mbox{\epsfig{figure=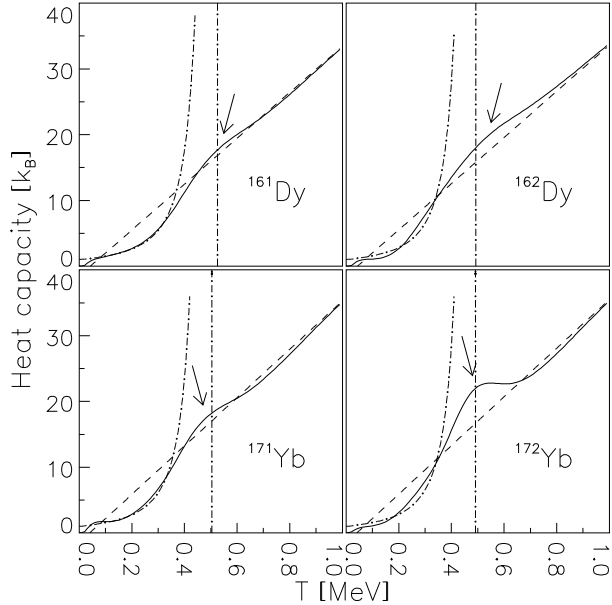,height=10cm}}
\caption{Semi-experimental heat capacity as function of temperature in the 
canonical ensemble for $^{161,162}$Dy and $^{171,172}$Yb. The dashed lines 
describe the approximate Fermi-gas heat capacity. The arrows indicate the first
local maxima of the experimental curve relative to the Fermi-gas estimates. The
dashed-dotted lines describe estimates according to a constant-temperature
level density formula, where $T$ is set equal to the critical temperature 
$T_c$ (vertical lines).}
\label{fig:4}
\end{figure}

Due to this strong smoothing over a huge range of excitation energies, one does
not expect to see fine structures in the canonical ensemble. This is clearly 
demonstrated in Fig.~\ref{fig:3}, where the canonical caloric curve is smooth
and the breaking of individual Cooper pairs is completely washed out. However,
the quenching of pairing correlations is manifested in the canonical 
heat-capacity curves (see Fig.~\ref{fig:4}). Deviating from a Fermi-gas 
estimate, the heat-capacity curves show pronounced S-shapes with local maxima
relative to the smooth Fermi-gas estimate. This behavior can be explained by
the fact that the level density exhibits a constant-temperature part at low
excitation energies. Therefore the canonical heat capacity curve 
$C_V=(1-T/\tau)^{-2}$ for a constant-temperature level density 
$\rho=C\exp(E/\tau)$ has been fitted to the data at low temperatures, and the
parameter $\tau$ is interpreted as the critical temperature for the quenching
of pairing correlations \cite{SB99}. The resulting critical temperatures are
given as horizontal and vertical lines in Figs.~\ref{fig:3} and \ref{fig:4}
respectively. We interpret the S-shape of the heat capacity as a fingerprint of
a second-order phase-transition-like phenomenon in finite systems, where the
transition goes from a phase with strong pairing correlations (usually referred
to as a superfluid phase) to a phase with weak pairing correlations (normal
fluid phase). This phase-transition-like phenomenon has been anticipated by
many theoretical works \cite{SY63,NA97,DK95,RH98}.

\section{Radiative strength function}
\label{sec:4}

\begin{figure}[htb!]\centering
\mbox{\epsfig{figure=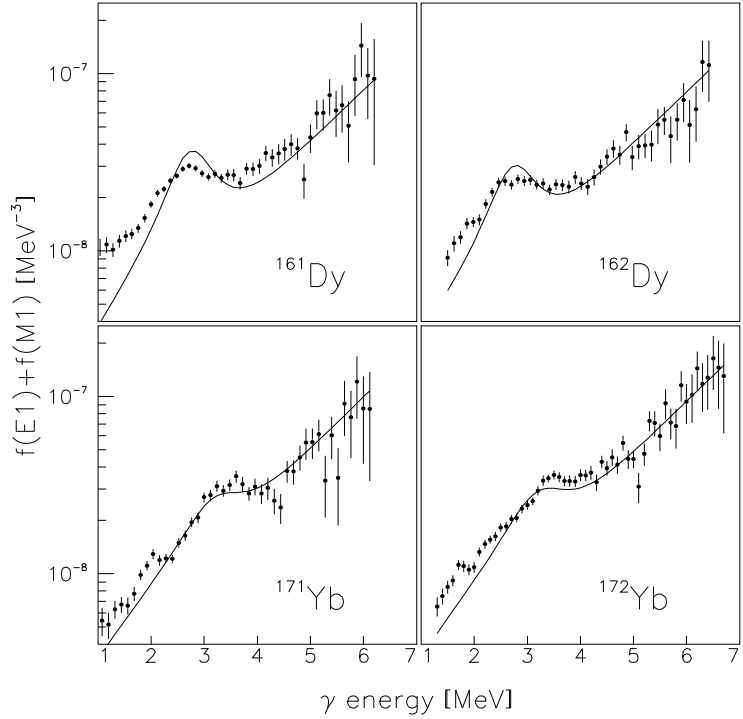,height=12.1cm}}
\caption{Radiative strength function of $^{161,162}$Dy and $^{171,172}$Yb (data
points). The absolute normalization of the data is still preliminary. The error
bars show the experimental uncertainties only. The solid lines are strength 
function models (see text) where all parameters are taken from other 
experimental systematics and nothing was fitted to our data beside $k\sigma_p$
(see text).}
\label{fig:5}
\end{figure}

Fig.~\ref{fig:5} shows the radiative strength functions of $^{161,162}$Dy and 
$^{171,172}$Yb compared to model calculations. For the theoretical calculation,
we have used the E1 model of Sirotkin \cite{Si86}, where we take the expression
for the temperature-dependent width of Kadmenski{\u{\i}} et al.\ \cite{KM83}.
The parameters are taken from an interpolation of the experimental systematics 
of \cite{GL81}. The temperature has been assumed as constant with 
$T\sim 500$~keV. For the M1 model we simply take a Lorentzian, where the 
parameters for the centroids and widths are taken from \cite{KU90} and the 
parameters for the resonance strengths are taken from $f_{M1}/f_{E1}$ 
systematics \cite{KU94}, evaluated at $E_\gamma=B_n-1$~MeV. For the pygmy 
resonance, we use again a Lorentzian with parameters from an interpolation  
based on the experimental systematics \cite{IK86}. It is amazing that the model
calculation can fit our data so well. Both, the absolute value as well as the 
slope of the experimental strength functions could be reproduced without 
fitting any parameter from the models except $k\sigma_p$. Here we had to reduce
the parameters for the pygmy resonance strength $k\sigma_p$ by 30--70\% as the 
only compromise to our data. 

In the following, we want to investigate the strength of the pygmy resonance, 
which is the only parameter we had to fit in order to describe our data. For
this reason, we divide our primary $\gamma$-ray matrix into four subsets of
distinct excitation energy bins. Each excitation energy bin is 1~MeV broad, 
thus we can assume that the nuclear temperature within every excitation energy 
bin is constant and the Brink-Axel hypothesis remains valid. However, for the
different excitation energy bins the nuclear temperature is in general 
different. We extract radiative strength functions from those four excitation 
energy bins. In that way we obtain radiative strength functions for four 
different nuclear temperatures. This provides an easy way to investigate the
temperature dependence of the radiative strength function. 

\begin{figure}[htb!]\centering
\mbox{\epsfig{figure=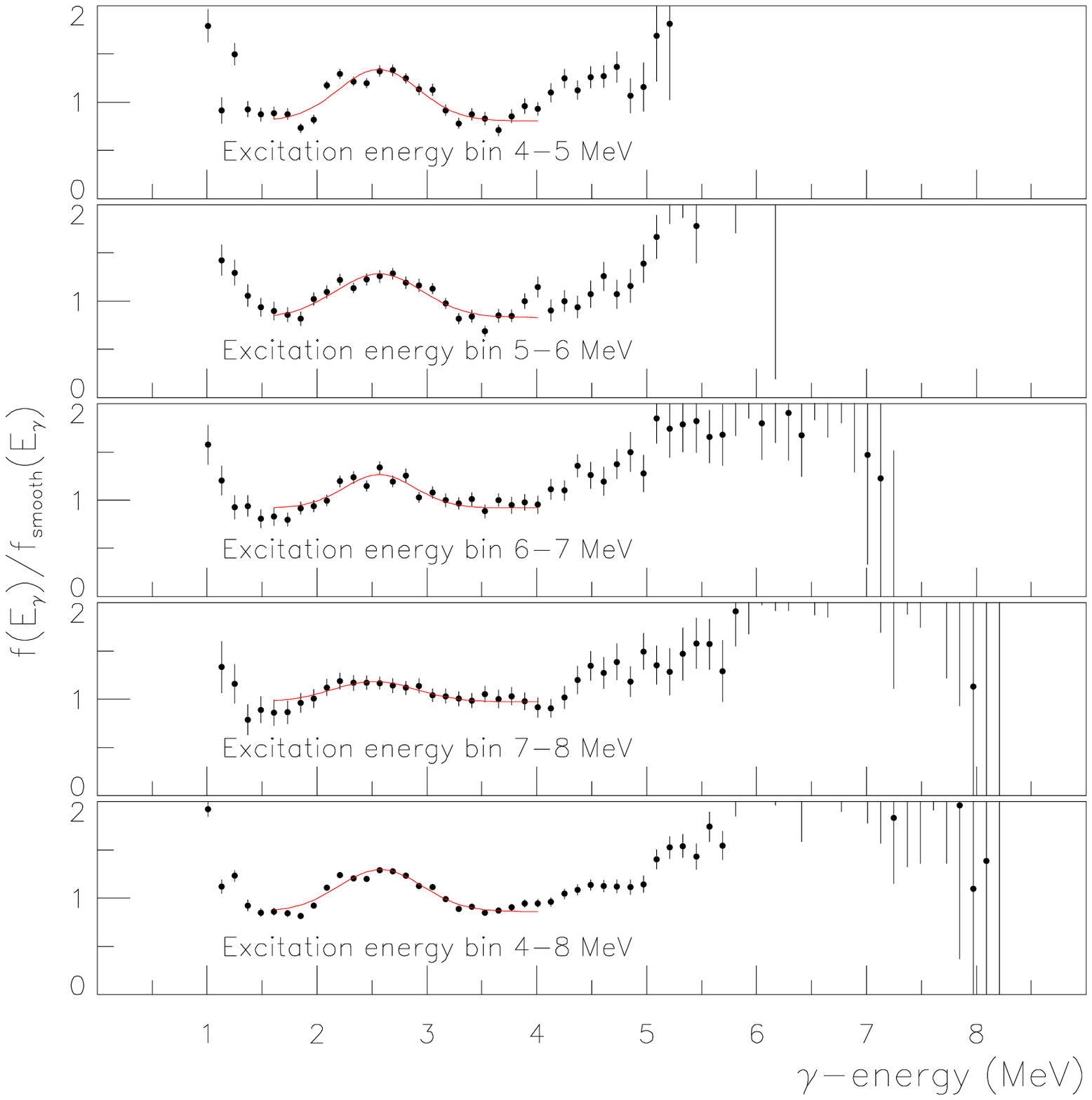,height=12.1cm}}
\caption{Relative radiative strength functions extracted from different 
excitation energy bins. The data here and in the following discussion are from 
the $^{162}$Dy($^3$He,$^3$He')$^{162}$Dy reaction. The radiative strength 
functions are all divided by the same smooth strength function $CE_\gamma^n$ 
with $n\approx 1.2$ in order to enhance the pygmy resonance structure.}
\label{fig:6}
\end{figure}

\begin{figure}[htb!]\centering
\mbox{\epsfig{figure=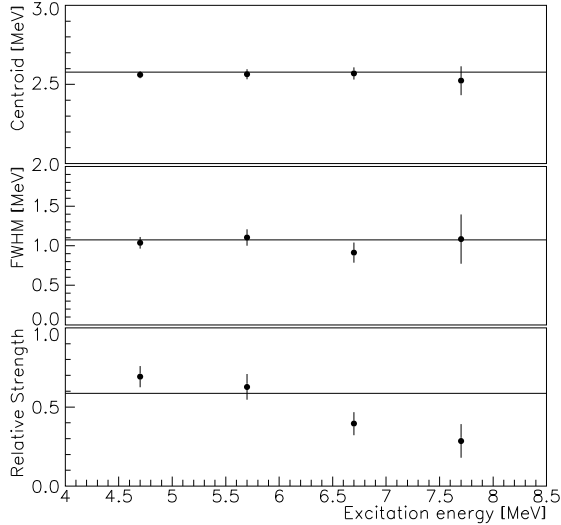,height=8cm}}
\caption{Fit parameters from a Gaussian fit to the pygmy resonance structure
(see Fig.~\protect{\ref{fig:6}} for different excitation energy bins. The 
centroid and the width are constant and fit nicely into the systematics of
Ref.~\protect{\cite{IK86}}. The relative strength is decreasing with increasing
excitation energy. The lines indicate average values, taking into account the 
whole data set.}
\label{fig:7}
\end{figure}

\begin{figure}[htb!]\centering
\mbox{\epsfig{figure=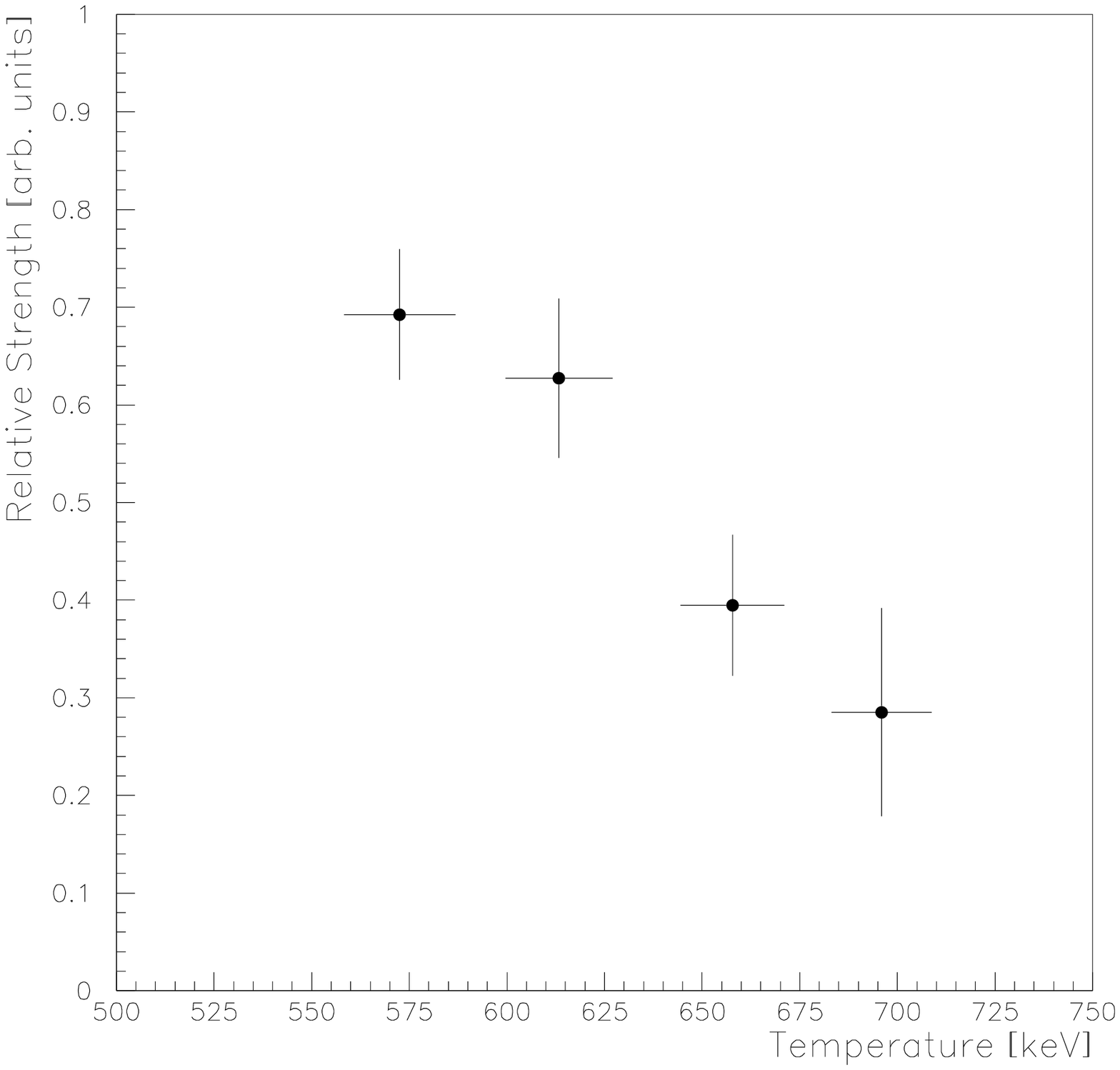,height=8cm}}
\caption{Temperature dependence of the relative pygmy resonance strength.}
\label{fig:8}
\end{figure}

In Fig.~\ref{fig:6}, the relative radiative strength functions from the 
different excitation energy bins are plotted. One can see immediately that the 
pygmy resonance strength is highly temperature dependent. On the other hand, 
the general slope of the radiative strength function is constant, thus the 
gross features of the strength function are rather independent of nuclear 
temperature, justifying the use of the Brink-Axel hypothesis for our data 
analysis. The fit parameters of the pygmy resonance for the different 
excitation energy bins are shown in Fig.~\ref{fig:7}. Obviously, only the 
resonance strength of the pygmy resonance shows a pronounced temperature 
dependence, whereas the centroid and the width are nearly independent of 
temperature. The temperature dependence becomes even much more obvious when we 
actually translate the excitation energy bins into nuclear temperature using 
the canonical caloric curve of Fig.~\ref{fig:3}. The result is given in 
Fig.~\ref{fig:8}, where the temperature dependence of the pygmy resonance 
strength is shown. A clear quenching of the pygmy resonance strength as 
function of temperature is observed.

We have to speculate on the physical origin of the observed quenching. In the
first place, it is not at all clear if the pygmy resonance is a phenomenon in
the electric or magnetic dipole strength function. Igashira et al.\ \cite{IK86}
favor electric dipole strength without measuring the parity of the transition,
whereas in other works, spin-flip \cite{GR84} or orbital \cite{WB88} (scissors 
mode) M1 strength has been proposed. Anyhow, we might assume a strong 
dependence of the pygmy resonance strength on the deformation parameter 
$\delta$ (like it is observed for the scissors mode \cite{PB98}). A quenching
of the pygmy resonance strength would then correspond to a shape transition
of the nucleus from deformed to spherical. This temperature-induced shape
transition was indeed anticipated for $^{170}$Dy in \cite{KD97} at temperatures
around 500~keV. Therefore, we speculatively interpret the quenching of the 
pygmy resonance strength as a fingerprint for a temperature-induced shape 
transition. 

\section{Conclusion}
\label{sec:5}

A method to extract simultaneously level density and radiative strength 
function from primary $\gamma$-ray spectra without assuming any model for 
either of them has been presented. Thermodynamical quantities have been deduced
within the microcanonical and the canonical ensemble. We observe structures in 
these quantities which can be interpreted as breaking of Cooper pairs and 
quenching of pairing correlations and we observe a fingerprint of the 
phase-transition-like phenomenon from a superfluid-like phase to a 
normal-fluid-like phase. Further on, the critical temperature of this 
transition has been determined. We are able to reproduce our experimental 
strength functions by the use of models, where all parameters except the pygmy 
resonance strength are taken from other experimental systematics. The 
temperature dependence of the pygmy resonance strength has been investigated 
and a significant quenching around $T\approx 500$~keV has been observed, which 
we interpret tentatively as the result of a temperature-induced shape 
transition. 

\section{Acknowledgments}

The authors are grateful to E. A.~Olsen and J.~Wikne for providing the 
excellent experimental conditions. We thank A.~Voinov for many interesting 
discussions. We wish to acknowledge the support from the Norwegian Research 
Council (NFR).

\end{document}